\documentclass[aps,twocolumn,floats,prd,nofootinbib,,superscriptaddress,letterpaper]{revtex4} 
\usepackage[dvips]{graphicx} %
\usepackage{epsfig,amsmath}
\usepackage{amssymb}
\usepackage{color,comment}
\usepackage{bm}

\DeclareFontFamily{OT1}{pzc}{}
\DeclareFontShape{OT1}{pzc}{m}{it}%
            {<-> s * [1.10] pzcmi7t}{}
\DeclareMathAlphabet{\mathscr}{OT1}{pzc}%
                                {m}{it}

\newcommand{\be}{\begin{eqnarray} }
\newcommand{\ee}{\end{eqnarray} }
\newcommand{\bs}{\begin{split} }
\newcommand{\es}{\end{split} }

\newcommand{\Mpl}{M_{\mathrm Pl}}

\newcommand{\tN}{\tilde N}
\newcommand{\e}{\epsilon}

\renewcommand{\comment}[1]{}

\def\[{\left [}
\def\]{\right ]}
\def\({\left (}
\def\){\right )}
\def\p{\partial}

\def\r2{\sqrt{2}}

\def\drs{\Delta_R^2}

\begin{document}

\title{On the Three Primordial Numbers}

\author{Roberto Gobbetti}
\affiliation{Institute for Theoretical Physics and Center for Extreme Matter and Emergent Phenomena,
Utrecht University, Leuvenlaan 4, 3584 CE Utrecht, The Netherlands}

\author{Enrico Pajer}
\affiliation{Institute for Theoretical Physics and Center for Extreme Matter and Emergent Phenomena,
Utrecht University, Leuvenlaan 4, 3584 CE Utrecht, The Netherlands}

\author{Diederik Roest}
\affiliation{Van Swinderen Institute for Particle Physics and Gravity, University of Groningen, \\ Nijenborgh 4, 9747 AG Groningen, The Netherlands}

\begin{abstract}
Cosmological observations have provided us with the measurement of just three numbers that characterize the very early universe: $ 1-n_{s} $, $ N $ and $ \ln \drs$. Although each of the three numbers individually carries limited information about the physics of inflation, one may hope to extract non-trivial information from relations among them.
 Invoking minimality, namely the absence of ad hoc large numbers, we find two viable and mutually exclusive inflationary scenarios. The first is the well-known inverse relation between $1- n_{s} $ and $ N $. The second implies a new relation between $ 1-n_{s} $ and $  \ln \drs $, which might provide us with a handle on the beginning of inflation and predicts the intriguing \textit{lower} bound on the tensor-to-scalar ratio $ r> 0.006 $ ($ 95\% $ CL).

\end{abstract}
\date{\today}

\maketitle

 
\section{Introduction}

Half a century after its discovery in 1965, one would like to take stock of what we have learned from the Cosmic Microwave Background (CMB), and what this implies for our understanding of the primordial Universe. While an optimist might marvel that there are (many) inflationary models giving rise to the required CMB measurements, a pessimist might complain that this stems from the small amount of data that needs to be explained. Indeed, notwithstanding the highly accurate measurements of the CMB, the amount of information on our primordial Universe today consists of many upper and lower bounds, but just \textit{three measured numbers}.

The first number is directly related to the amplitude of CMB temperature fluctuations at the level of one part in $10^5$. This translates into a power spectrum of scalar perturbations generated during inflation that takes the value \cite{COBE, Ade:2015lrj} 
 \begin{align}\label{amplitude}
 	-\ln\drs (k_{\ast}) = 19.932 \pm 0.034\quad \mathrm{(68\%\,CL)}\,,
 \end{align}
at the pivot scale $ k_{\ast}=0.05 \,\mathrm{Mpc^{-1}} $. Throughout this paper we will always consider the logarithm of $ \drs $ rather than $ \drs $ itself, because it is precisely the former that is naturally related to the second primordial number, namely the scalar spectral index \cite{Ade:2015lrj}
\be \label{tilt}
	1-n_{s}(k_{\ast})&\equiv &- \frac{\partial \ln \Delta_{R}^{2} (k_{\ast})}{\partial  \ln k}\\
	& =& 0.0355 \pm 0.0049 \quad \mathrm{(68\%\,CL)}\,. \nonumber
\ee
The scale invariant, Harrison-Zel'dovich power spectrum with $n_s=1$ is therefore conclusively ruled out. Instead we have emerged from a primordial Universe with a percent-level red tilt, $n_s <1$.

Fluctuations at the pivot scale $ k_{\ast} $ left their sound horizon a number $N=N_{\ast}$ of efolds before the end of inflation (in our conventions, $N$ decreases in time). This number can be extracted from the expansion history of the universe that we infer from our measurements of matter, Dark Energy and radiation abundance\footnote{For example, for a toy universe with only radiation, $ N_{\ast} $ is equal to the number of efolds of \textit{decelerated} expansion since the end of inflation to today, which depends only logarithmically on the unknown scale of reheating.}. It is therefore an observable, not unlike $ 1-n_{s} $ or $ \ln\drs $. It is true however that the uncertainty on this number is much bigger than on the other two, due to the lack of constraints on the history of our universe between reheating and Big bang nucleosynthesis. What matters for the purpose of this paper is that $ N_{\ast} $ is a number roughly between 40 and 60. This uncertainty will change somewhat the predictions of the model in section \ref{2} but not its conclusions. Remarkably, the new model presented in section \ref{3} is independent of $N_{\ast} $, and therefore bypasses this uncertainty completely.

While it is important to continue our effort to extract additional information about the primordial universe, for example from non-Gaussianity or tensor modes, we should also take the time to ponder what we have learned from the three primordial numbers we \textit{have} measured so far. In particular, in this paper we will ask if current data are already accurate enough to rule out or perhaps suggest simple relations among $ N $, $ 1-n_{s} $ and $ \ln \drs $. 

Note that it is hard to extract information about the mechanism of inflation from the value of each one of these three numbers taken individually. The reason is that the values we have measured depend on when we (i.e.~the observers) happen to live in the history of the universe.

To make this point sharper, imagine a universe without Dark Energy and a civilization that first measures the CMB hundreds of Hubble times in the future from now. Each of the three primordial numbers will most probably take a different value. To proceed from there one would then need to speculate about the nature of observers and how they are distributed in time, pushing deeper into uncharted territory. To overcome this difficulty, we make one choice and one assumption. The choice is to consider only relations among at least two of the three primordial numbers. \textit{The assumption is that these relations should be valid over a wide range of efolds}, and not just within the window of about 7 efolds that we have been able to probe with cosmological observations.

The assumption is motivated by the fact that inflation provides a very robust and elegant explanation of the approximate scale invariance of primordial perturbations in terms of the isometries of quasi-de Sitter spacetime. It is therefore enticing to interpret observations within this elegant paradigm rather than believe that they crucially rely on the specific time-scale of intelligent, carbon-based life forms in a Dark Energy dominated universe.

We take \textit{minimality} as our guiding principle: a model should have as few unexplained hierarchies as possible. More specifically, we do not allow ourselves to introduce any ad hoc very large or very small number. Instead we ask whether models exist in which the smallness of $ 1-n_{s} $ and the largeness of $ N_{\ast} $ or $ -\ln \drs $ can be directly related to each other. Of course there are many models that fit the data that are \textit{not} minimal in this specific sense. Whether our own universe is described by one of those models or by a minimal one should be decided on the basis of observations. In the following we will construct two models that are minimal in this sense and show that both make testable predictions and could therefore be ruled out in the not so distant future. 

\begin{figure}[t]
\centering
\includegraphics[width=.35\textwidth]{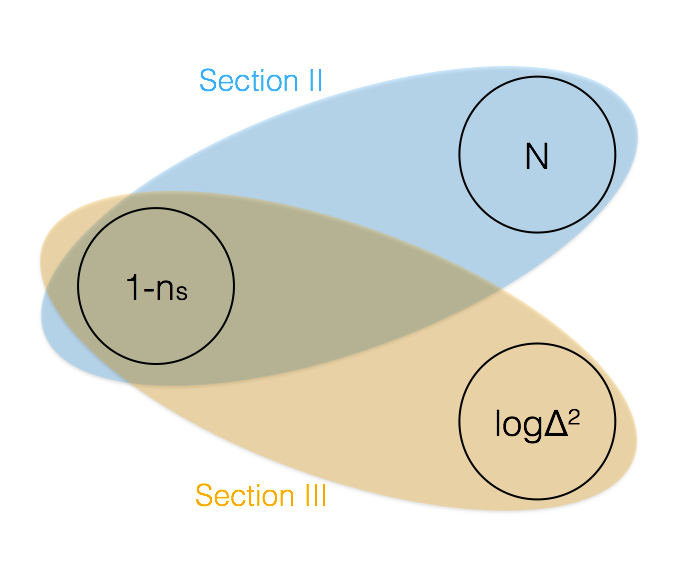}
\caption{\it Only two pairs of the three measured primordial numbers can be related in a minimal way. The two relations involve the spectral index and either the number of efolds (section \ref{2}) or the amplitude (section \ref{3}). The two options are mutually exclusive.\label{fig}}
\vspace{-0.5cm}
\end{figure}

Our first result is that \textit{only two of the three primordial numbers can be related in a minimal way}, but not all three of them at the same time (see figure \ref{fig}). In other words, two numbers must be fixed by observation. Since we measure a \textit{red} spectral tilt, it must be that $ -\ln \drs $ becomes an even larger number at shorter scales, whereas $ N $ decreases towards an order one number. This strongly suggests that the measured largeness of $ -\ln \drs(k_{\ast}) $ is not related to the largeness of $ N_{\ast} $, proving our initial point. There are therefore only two mutually exclusive options: the smallness of $ 1-n_{s}(k_{\ast}) $ is related either to the largeness of $ N_{\ast} $ or to the largeness of $ -\ln \drs(k_{\ast}) $. We will consider each possibility in turn in the next two subsections and conclude with some general comments on the role played by the running of the spectral tilt in this distinction.

\section{Tilt and efolds}\label{2}

We follow the approach proposed in \cite{Mukhanov, Roest:2013fha} (see \cite{Huang} for an early discussion) to infer a relation between the tilt of the power spectrum and the number of efolds. We assume that the relation
 \be \label{xandN}
  1-n_{s}(k_{\ast}) = \frac{\gamma}{N_{\ast}}\,,
 \ee
which we \textit{observe} at CMB scales with $\gamma$ an order one constant, holds for a much larger range of scales and hence range of $N$. The latest Planck measurement of the spectral index yields
\begin{align} \label{gamma}
	\gamma = (2.1 \pm 0.3)\times \frac{N_{\ast}}{60} \quad \mathrm{(68\%\,CL)} \,.
\end{align} 
It would be interesting to understand if $ \gamma $ being an integer has some special implication for the (non-linear) regime of primordial perturbations. Here we simply notice that the uncertainties on $ 1-n_{s}(k_{\ast}) $ and $ N_{\ast} $ are so large that the data do not yet provide any indication that this holds for our universe.

The Ansatz in \eqref{xandN} determines the entire scale dependence of the primordial spectrum of scalar perturbations. In particular, now we can turn the definition of the tilt \eqref{tilt} into a differential equation for $\drs(N)$ using $ dN=-d\ln k $. The solution is
 \begin{align}
 -\ln \drs (N) = C -\gamma \ln N \,,
 \end{align}
 where the integration constant is fixed by measurements to $ C=-\ln \drs(k_{\ast})+\gamma \ln N_{\ast}\simeq 29$. This relation is not minimal because it requires introducing the large number $ C \gg 1$. This confirms our general argument at the end of the introduction that the negative sign of the spectral tilt prevents a minimal relation between the largeness of $ N $ and $ -\ln \drs $. In this class of models, like in almost all other models of inflation in literature, the normalization of the power spectrum (a.k.a.~the COBE normalization) is fixed ad hoc, by appropriately choosing the scale of inflation. A very different perspective will be proposed in the next section.

An interesting consequence of the $1/N$ relation \eqref{xandN} is that one can determine the tensor-to-scalar ratio $r$ from the slow-roll relations:
\be
  1 - n_{s}& =  &2\epsilon-\frac{\epsilon_{,N}}{\epsilon}-\frac{c_{s,N}}{c_{s}}  \label{eom} \\
     &  = &\frac{r}{8 c_s} - \frac{1}{r}\frac{\p r}{\p N}  \label{eom2} \,, 
\ee
where we allowed for a time-dependent speed of sound $c_s(N)$ and used $r = 16 c_s \epsilon$. The general solution of \eqref{eom2} is
\be
	r(N)=8\left[  N^{\gamma} \left(  A -\int_{N_{0}}^{N} \frac{dn }{n^{\gamma} c_{s}(n) } \right) \right]^{-1}\,.
\ee
When the speed of sound $c_s=1$ we find the result \cite{Creminelli:2014nqa}
\be\label{eofN}
	r = \frac{16 (\gamma -1)}{2N + A(\gamma-1)N^{\gamma}} \,,
\ee
where $A$ is an integration constant. This solution has two regimes corresponding to $A\simeq0$ or $A\simeq 1$ (excluding the possibility that we are observing the transition between these two, as this requires a fine-tuned value for $A$ which would invalidate the premise of this work): 
\begin{itemize}
\item 
The first regime $A \simeq 0$ leads to the same predictions as monomial inflation:
 \be \label{quadratic}
  r \simeq \left( 0.15 \pm 0.04 \right)\times \frac{60}{N_{\ast}} \quad \mathrm{(68\%\,CL)}\,.
  \ee
This is already in strong tension with CMB observations that put the $95\% $ CL at $ r\leq 0.08 $ (eq (167) of \cite{Ade:2015lrj}). 
\item
 The second regime $A\simeq 1 $ implies {$r\sim 16/N^\gamma$} and hence is typically at the permille level. In this case the exact value is not fixed due to the integration constant $ A $ and the uncertainty on $ N_{\ast} $. 
\end{itemize}
To gain insight into the more general case of a non-trivial speed of sound \cite{Zavala:2014bda}, it is instructive to look at \eqref{eom}. Two distinct cases can be discerned:
\begin{itemize}
  \item
$c_s$ is constant or slowly varying, but possibly different from one. Then the last term in \eqref{eom} drops out and the analysis is the same as for $c_s=1$ except that $r = 16 c_s \epsilon$ is smaller by a factor of $ c_{s} $. This allows one to change the overall normalization of the tensor spectrum at the expense of having a non-unity speed of sound, and hence a non-Gaussianity of order $f_{\rm NL}^{eq,or} \simeq 1-1/c_s^2$ \cite{Cheung:2007st} or larger because of $ c_{3} $ \cite{Senatore:2009gt}. Therefore one can only significantly alter the prediction \eqref{quadratic} at the price of having a sizable non-Gaussianity.
 \item
$c_s$ has a significant $N$-dependence. Then either the last term in \eqref{eom} dominates, or all three terms contribute at the same order. In the former case one has
\begin{align}
	c_s = c_s (N=1) N^{-\gamma} \,,
\end{align}
which for  $\gamma\sim2$ (see \eqref{gamma}) yields too much variation in the speed of sound between $N=60$ and $N=1$. This would imply either a violation of the constraints on non-Gaussianity at $N=60$ or a superluminal speed of sound as $N$ approaches small values. We are left only with the possibility that all terms contribute at the same order, leading to a scaling of $c_s$ anywhere in between $N^0$ and $N^{-\gamma}$; in fact, by the same arguments as above, one would require this scaling to be fairly close to a constant.
 \end{itemize}

The models in this section share some common traits. First, as the slow-roll parameters scale as $1/N$, 
the universe asymptotes a de Sitter spacetime in the far past (when $N$ is large). It can be shown \cite{Mukhanov:2014uwa} that there is an eternal inflation regime at some point in the past. Second, the running of the spectral index is quadratic in the slow-roll parameters and negative for all of these models (see section \ref{4}). Third, a large number of inflationary models with specific scalar potentials falls in this class, ranging from chaotic inflation with monomial potentials $V \simeq \phi^\lambda$ to e.g.~Starobinsky inflation with $\gamma=2$. The field range of these models can be analyzed universally, allowing for a generalization of the Lyth bound  \cite{Lyth, Garcia-Bellido:2014wfa}.

 
\section{Amplitude and tilt}\label{3}

As we have discussed, the redness of the spectral tilt implies that $ -\ln \drs $ grows as $ N $ decreases, meaning that the two measured large numbers $ -\ln \drs(k_{\ast}) $ and $ N_{\ast} $ cannot be minimally related. This suggests that $ -\ln \drs $ might instead be related to $\tN\equiv N_0 - N$, with $ N_{0} $ some fixed time in our past. In words this means that the largeness of $ -\ln \drs $ (namely the distance from some eternal inflation regime $-\ln \drs \simeq 1 $) might be related to the largeness of the number of efolds $ \tN $ elapsed from some initial event in our past at $ N=N_{0} $. Naively this seems a cul-de-sac: we measure $  -\ln \drs (k_{\ast})$ and $ N_{\ast} $ but know nothing about $ N_{0} $ and so we cannot use observations to guess the relation $ -\ln\drs(\tN) $, as we did in the previous section for $ 1-n_{s}(N) $. 

An interesting way out can be found as follows. If we assume that $ \tN $ is the relevant large number in the problem, such that $ -\ln\drs= -\ln\drs(\tN) $ and $ 1-n_{s}=1-n_{s}(\tN) $, then we can invert the last relation, namely $ \tN=\tN(1-n_{s}) $ and substitute it into the first. This allows us to bypass $ \tN $ and look for a minimal relation between $ -\ln \drs $ and $ 1-n_{s} $. The measured valued of these two quantities at cosmological scales suggest the minimal relation
\begin{align} \label{deltaandx}
	-\ln(\drs) = \frac{\lambda}{1-n_s} \,,
\end{align}
with $\lambda$ a constant fixed by the measurements \eqref{amplitude} and \eqref{tilt}
\begin{align}
	\lambda = 0.7 \pm 0.1 \quad \mathrm{(68\%\,CL)}\,,
\end{align}
which is indeed order one as required by minimality. The Ansatz \eqref{deltaandx} and the definition of the tilt \eqref{tilt} lead to a differential equation for the tilt that can be solved to give
 \be \label{nsmod2}
 	(1-n_s)^2 = \frac{\lambda}{2(N_0-N)}=\frac{\lambda}{2\tN}\,,
 \ee
 where the initial time $N_0$ appears naturally as an integration constant needed to offset the possibly negative denominator. This confirms our expectations that $\tN \equiv N_0 - N$ is the relevant large number in the problem. The measurement of the tilt at the pivot scale fixes the corresponding value
 \be
 	\tN_{\ast} = 278 \pm 34 \quad \mathrm{(68\%\,CL)}\, .
 \ee
 Using $ N_{\ast}=60 $, the integration constant is completely determined: $N_ 0 \simeq 338$.

The $N$-dependence of the amplitude of the power spectrum is easily read off from \eqref{deltaandx}:
 \begin{align}
- \ln   \drs = \sqrt{2 \lambda \tilde N} \,,
 \end{align}
 which indeed relates the largeness of $ -\ln \drs $ to that of $ \tN $, as expected.
$ \e $ can be computed from \eqref{eom} assuming for simplicity a constant speed of sound
  \be \label{r2}
     \epsilon = \frac{1/2}{\sqrt{2 \lambda \tN} - 1 + 8 A e^{- \sqrt{2 \lambda \tN}}} 
     &\simeq& \frac{  1-n_{s}}{2}\,,
  \ee
where in the last line we neglected the exponentially suppressed term proportional to the integration constant $ A $ and  assumed $ \tN\gg 1 $. Unlike for the models in the previous section, here the prediction for $ r $ is \textit{independent} of $ N_{\ast} $, hence bypassing our ignorance of the details of reheating. Neglecting the exponentially suppressed term in \eqref{r2} (but not the -1), we find
\be
	r&=&16\, \epsilon\, c_s^{1+2\epsilon} \label{new_rel}\simeq 0.30\times  c_{s}^{1.0355}\,.
\ee
Here, we improved the usual leading order relation $ r=16 \e c_{s} $ with the slow-roll suppressed correction discussed in \cite{Baumann:2014cja} (leading to changes of order $ 10\% $ or less), and used $2\epsilon\simeq1-n_s\simeq 0.0355$.

The relation between the spectral index and the tensor-to-scalar ratio\footnote{For a similar analysis of non-Gaussianities when the tensor-to-scalar ratio is large see \cite{D'Amico:2014cya}.} is plotted in the upper panel of figure 2 for various values of $ c_{s} $. Clearly $ c_{s}=1 $ is excluded. From the Planck plus Bicep constraint $ r<0.08 $ at $  95\%$ CL \cite{Ade:2015lrj}, we find the upper bound\footnote{This model therefore predicts new physics beyond the standard, weakly-coupled, slow-roll paradigm in the sense of \cite{Baumann:2014cja}.} $c_s < 0.28$, implying equilateral or orthogonal non-Gaussianity of order one or larger. The lower panel of figure 2 shows the size of equilateral and orthogonal non-Gaussianities for the same values of $c_s$. Following the analysis of \cite{Senatore:2009gt}, the marginalized constraint is $c_s > 0.024$ at $ 95\% $ CL \cite{Ade:2015ava}. This implies a \textit{lower} bound on the tensor-to-scalar ratio: $ r>0.006 $ at $ 95\% $ CL ($ r>0.008 $ at $ 68\% $ CL). Therefore, within the theoretical priors of this model,  {\it a vanishing tensor-to-scalar ratio is excluded by the constraints on non-Gaussianity}. 

\begin{figure}[t]
\centering
\includegraphics[width=.47\textwidth]{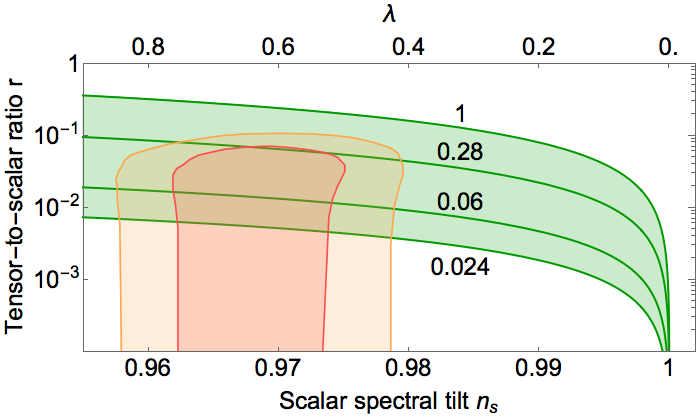}
\includegraphics[width=.47\textwidth]{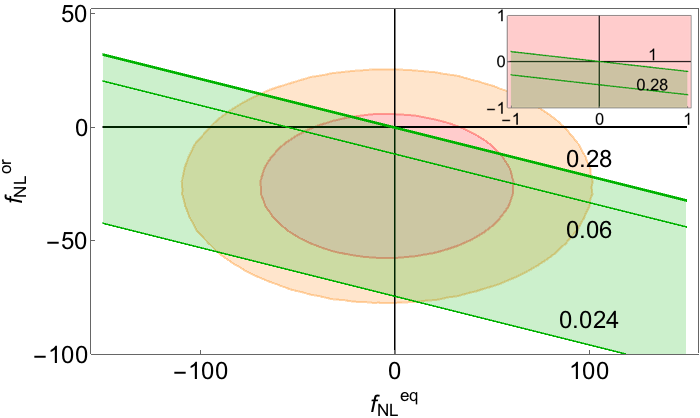}
\caption{\it We show the $68\%$ and $95\%$ CL contours for the spectral index and tensor-to-scalar ratio (upper panel) and non-Gaussianity (lower panel)  \cite{Ade:2015lrj, Ade:2015ava}. In both panels we superimpose the specific predictions of this model for various values of $c_s = (0.024, 0.06,  0.28, 1)$. \label{fig2}}
\vspace{-0.5cm}
\end{figure}

Finally, we can integrate $\epsilon = r/(16c_{s})$ to find the scalar potential that corresponds to this model (always for constant $c_s$). Using again the approximated form of \eqref{r2}, as appropriate for $\tN\gg 1$, we find
\be
	V \simeq V_0 \exp \left[  - (3 \sqrt{\lambda} \phi / (2\Mpl))^{2/3} \right]  \, ,
\ee
which is valid for $ \phi\gg \Mpl $.

Two comments are in order. First, as we move back in time approaching $ N=N_{0} $ three things happen almost simultaneously that invalidate our analysis: the energy density becomes of order $ \Mpl^{4} $, the slow-roll parameters become of order one and the power spectrum becomes of order one. Depending on the details of the UV-completion of gravity it is therefore plausible that the past of this model contains \textit{no regime of eternal inflation}. A similar approach has been advocated in \cite{Mukhanov:2014uwa}.
Second, the slow roll parameters decrease towards the end of inflation (see \eqref{nsmod2}). This implies that in order to end inflation we must introduce a waterfall field, whose corrections to our derivation are negligible at $ N\gtrsim 60 $ but grow large as $ N $ approaches zero (e.g.~a $ 1/N^{2} $ correction to $ 1-n_{s} $). Notice that the relation \eqref{deltaandx} and the prediction for $r  $ in \eqref{r2} are \textit{always} valid as long as the contribution from the waterfall field is negligible, irrespectively of when inflation ends.

 
\section{Discussion}\label{4}

The two models discussed in this Letter represent benchmarks of simplicity in an otherwise unlimited landscape of possibilities. By excluding them, we will learn, even in the absence of a detection, that our universe happens to be a little bit less minimal, at least in the specific sense discussed here. 

It is interesting to consider the running of the tilt, which is observationally constrained to  \cite{Ade:2015lrj}
\be \label{running}
	\alpha_s =  -\frac{d}{dN} n_s = - 0.0085 \pm 0.0076  \quad \mathrm{\left(  68\% \,CL\right)}\, .
\ee
In the model of section \ref{2}, $ \alpha $ is naturally \textit{second order} in $1-n_s$ and \textit{negative}:
\be\label{run1}
	\alpha_s = - \frac{\gamma}{N^2}=-\frac{1}{\gamma} \left(  1-n_{s}\right)^{2}\, .
\ee
At the pivot scale, the central value is $ \alpha_{s}(k_{\ast})=- 6\times 10^{-4} $. This permille and negative value turns out to be a generic prediction that applies to many other inflationary models \cite{Garcia-Bellido:2014gna}. In the model of section \ref{3}, $ \alpha $ is instead \textit{third order} in $1-n_s$ and \textit{positive}:
\be\label{run2}
	   \alpha_s = \sqrt{\frac{\lambda}{8 \tilde N^3}}=\frac{1}{\lambda} \left(  1-n_{s}\right)^{3}\, .
\ee
We find the central value $ \alpha_{s}(k_{\ast})= 6\times 10^{-5} $, an order of magnitude smaller than in the previous case. This is a much less common scaling than that in \eqref{run1}. A very interesting generalization of this fact with implications for the current generation of CMB polarization experiment will be discussed elsewhere. In both cases the running will be most likely undetectable with Euclid \cite{Amendola:2012ys} or CMB spectral distortion as measured e.g.~by PIXIE \cite{Kogut:2011xw}. A running at order  $ (1-n_{s})^{2} $ might be within reach once we colonize the dark side of the moon. Notice however, that given the \textit{lower} bound on $ r $, the model in section \ref{3} can be ruled out or confirmed in the next few years.

 
\section*{Acknowledgements}


We would like to thank Paolo Creminelli, Guido D'Amico, Matthew Kleban, Marjorie Schillo, Gabriele Trevisan, Matias Zaldarriaga and Ivonne Zavala for useful discussions and comments on the draft. E.P. is supported by the D-ITP consortium, a program of the Netherlands organization for scientific research (NWO) that is funded by the Dutch Ministry of Education, Culture and Science (OCW).

%
%

\bibliography{minimality}

\begin{thebibliography}{18}
\expandafter\ifx\csname natexlab\endcsname\relax\def\natexlab#1{#1}\fi
\expandafter\ifx\csname bibnamefont\endcsname\relax
  \def\bibnamefont#1{#1}\fi
\expandafter\ifx\csname bibfnamefont\endcsname\relax
  \def\bibfnamefont#1{#1}\fi
\expandafter\ifx\csname citenamefont\endcsname\relax
  \def\citenamefont#1{#1}\fi
\expandafter\ifx\csname url\endcsname\relax
  \def\url#1{\texttt{#1}}\fi
\expandafter\ifx\csname urlprefix\endcsname\relax\def\urlprefix{URL }\fi
\providecommand{\bibinfo}[2]{#2}
\providecommand{\eprint}[2][]{\url{#2}}

\bibitem[{\citenamefont{Smoot et~al.}(1992)\citenamefont{Smoot, Bennett, Kogut,
  Wright, Aymon et~al.}}]{COBE}
\bibinfo{author}{\bibfnamefont{G.~F.} \bibnamefont{Smoot}},
  \bibinfo{author}{\bibfnamefont{C.}~\bibnamefont{Bennett}},
  \bibinfo{author}{\bibfnamefont{A.}~\bibnamefont{Kogut}},
  \bibinfo{author}{\bibfnamefont{E.}~\bibnamefont{Wright}},
  \bibinfo{author}{\bibfnamefont{J.}~\bibnamefont{Aymon}},
  \bibnamefont{et~al.}, \bibinfo{journal}{Astrophys.J.}
  \textbf{\bibinfo{volume}{396}}, \bibinfo{pages}{L1} (\bibinfo{year}{1992}).

\bibitem[{\citenamefont{Ade et~al.}(2015{\natexlab{a}})}]{Ade:2015lrj}
\bibinfo{author}{\bibfnamefont{P.}~\bibnamefont{Ade}} \bibnamefont{et~al.}
  (\bibinfo{collaboration}{Planck}) (\bibinfo{year}{2015}{\natexlab{a}}),
  \eprint{1502.02114}.

\bibitem[{\citenamefont{Mukhanov}(2013)}]{Mukhanov}
\bibinfo{author}{\bibfnamefont{V.}~\bibnamefont{Mukhanov}},
  \bibinfo{journal}{Eur.Phys.J.} \textbf{\bibinfo{volume}{C73}},
  \bibinfo{pages}{2486} (\bibinfo{year}{2013}), \eprint{1303.3925}.

\bibitem[{\citenamefont{Roest}(2014)}]{Roest:2013fha}
\bibinfo{author}{\bibfnamefont{D.}~\bibnamefont{Roest}},
  \bibinfo{journal}{JCAP} \textbf{\bibinfo{volume}{1401}}, \bibinfo{pages}{007}
  (\bibinfo{year}{2014}), \eprint{1309.1285}.

\bibitem[{\citenamefont{Huang}(2007)}]{Huang}
\bibinfo{author}{\bibfnamefont{Q.-G.} \bibnamefont{Huang}},
  \bibinfo{journal}{Phys.Rev.} \textbf{\bibinfo{volume}{D76}},
  \bibinfo{pages}{061303} (\bibinfo{year}{2007}), \eprint{0706.2215}.

\bibitem[{\citenamefont{Creminelli et~al.}(2014)\citenamefont{Creminelli,
  Dubovsky, Nacir, Simonovic, Trevisan et~al.}}]{Creminelli:2014nqa}
\bibinfo{author}{\bibfnamefont{P.}~\bibnamefont{Creminelli}},
  \bibinfo{author}{\bibfnamefont{S.}~\bibnamefont{Dubovsky}},
  \bibinfo{author}{\bibfnamefont{D.~L.} \bibnamefont{Nacir}},
  \bibinfo{author}{\bibfnamefont{M.}~\bibnamefont{Simonovic}},
  \bibinfo{author}{\bibfnamefont{G.}~\bibnamefont{Trevisan}},
  \bibnamefont{et~al.} (\bibinfo{year}{2014}), \eprint{1412.0678}.

\bibitem[{\citenamefont{Zavala}(2015)}]{Zavala:2014bda}
\bibinfo{author}{\bibfnamefont{I.}~\bibnamefont{Zavala}},
  \bibinfo{journal}{Phys.Rev.} \textbf{\bibinfo{volume}{D91}},
  \bibinfo{pages}{063005} (\bibinfo{year}{2015}), \eprint{1412.3732}.

\bibitem[{\citenamefont{Cheung et~al.}(2008)\citenamefont{Cheung, Creminelli,
  Fitzpatrick, Kaplan, and Senatore}}]{Cheung:2007st}
\bibinfo{author}{\bibfnamefont{C.}~\bibnamefont{Cheung}},
  \bibinfo{author}{\bibfnamefont{P.}~\bibnamefont{Creminelli}},
  \bibinfo{author}{\bibfnamefont{A.~L.} \bibnamefont{Fitzpatrick}},
  \bibinfo{author}{\bibfnamefont{J.}~\bibnamefont{Kaplan}}, \bibnamefont{and}
  \bibinfo{author}{\bibfnamefont{L.}~\bibnamefont{Senatore}},
  \bibinfo{journal}{JHEP} \textbf{\bibinfo{volume}{0803}}, \bibinfo{pages}{014}
  (\bibinfo{year}{2008}), \eprint{0709.0293}.

\bibitem[{\citenamefont{Senatore et~al.}(2010)\citenamefont{Senatore, Smith,
  and Zaldarriaga}}]{Senatore:2009gt}
\bibinfo{author}{\bibfnamefont{L.}~\bibnamefont{Senatore}},
  \bibinfo{author}{\bibfnamefont{K.~M.} \bibnamefont{Smith}}, \bibnamefont{and}
  \bibinfo{author}{\bibfnamefont{M.}~\bibnamefont{Zaldarriaga}},
  \bibinfo{journal}{JCAP} \textbf{\bibinfo{volume}{1001}}, \bibinfo{pages}{028}
  (\bibinfo{year}{2010}), \eprint{0905.3746}.

\bibitem[{\citenamefont{Mukhanov}(2015)}]{Mukhanov:2014uwa}
\bibinfo{author}{\bibfnamefont{V.}~\bibnamefont{Mukhanov}},
  \bibinfo{journal}{Fortsch.Phys.} \textbf{\bibinfo{volume}{63}},
  \bibinfo{pages}{36} (\bibinfo{year}{2015}), \eprint{1409.2335}.

\bibitem[{\citenamefont{Lyth}(1997)}]{Lyth}
\bibinfo{author}{\bibfnamefont{D.~H.} \bibnamefont{Lyth}},
  \bibinfo{journal}{Phys.Rev.Lett.} \textbf{\bibinfo{volume}{78}},
  \bibinfo{pages}{1861} (\bibinfo{year}{1997}), \eprint{hep-ph/9606387}.

\bibitem[{\citenamefont{Garcia-Bellido
  et~al.}(2014)\citenamefont{Garcia-Bellido, Roest, Scalisi, and
  Zavala}}]{Garcia-Bellido:2014wfa}
\bibinfo{author}{\bibfnamefont{J.}~\bibnamefont{Garcia-Bellido}},
  \bibinfo{author}{\bibfnamefont{D.}~\bibnamefont{Roest}},
  \bibinfo{author}{\bibfnamefont{M.}~\bibnamefont{Scalisi}}, \bibnamefont{and}
  \bibinfo{author}{\bibfnamefont{I.}~\bibnamefont{Zavala}},
  \bibinfo{journal}{Phys.Rev.} \textbf{\bibinfo{volume}{D90}},
  \bibinfo{pages}{123539} (\bibinfo{year}{2014}), \eprint{1408.6839}.

\bibitem[{\citenamefont{Baumann et~al.}(2015)\citenamefont{Baumann, Green, and
  Porto}}]{Baumann:2014cja}
\bibinfo{author}{\bibfnamefont{D.}~\bibnamefont{Baumann}},
  \bibinfo{author}{\bibfnamefont{D.}~\bibnamefont{Green}}, \bibnamefont{and}
  \bibinfo{author}{\bibfnamefont{R.~A.} \bibnamefont{Porto}},
  \bibinfo{journal}{JCAP} \textbf{\bibinfo{volume}{1501}}, \bibinfo{pages}{016}
  (\bibinfo{year}{2015}), \eprint{1407.2621}.

\bibitem[{\citenamefont{D'Amico and Kleban}(2014)}]{D'Amico:2014cya}
\bibinfo{author}{\bibfnamefont{G.}~\bibnamefont{D'Amico}} \bibnamefont{and}
  \bibinfo{author}{\bibfnamefont{M.}~\bibnamefont{Kleban}},
  \bibinfo{journal}{Phys.Rev.Lett.} \textbf{\bibinfo{volume}{113}},
  \bibinfo{pages}{081301} (\bibinfo{year}{2014}), \eprint{1404.6478}.

\bibitem[{\citenamefont{Ade et~al.}(2015{\natexlab{b}})}]{Ade:2015ava}
\bibinfo{author}{\bibfnamefont{P.}~\bibnamefont{Ade}} \bibnamefont{et~al.}
  (\bibinfo{collaboration}{Planck}) (\bibinfo{year}{2015}{\natexlab{b}}),
  \eprint{1502.01592}.

\bibitem[{\citenamefont{Garcia-Bellido and
  Roest}(2014)}]{Garcia-Bellido:2014gna}
\bibinfo{author}{\bibfnamefont{J.}~\bibnamefont{Garcia-Bellido}}
  \bibnamefont{and} \bibinfo{author}{\bibfnamefont{D.}~\bibnamefont{Roest}},
  \bibinfo{journal}{Phys.Rev.} \textbf{\bibinfo{volume}{D89}},
  \bibinfo{pages}{103527} (\bibinfo{year}{2014}), \eprint{1402.2059}.

\bibitem[{\citenamefont{Amendola et~al.}(2013)}]{Amendola:2012ys}
\bibinfo{author}{\bibfnamefont{L.}~\bibnamefont{Amendola}} \bibnamefont{et~al.}
  (\bibinfo{collaboration}{Euclid Theory Working Group}),
  \bibinfo{journal}{Living Rev.Rel.} \textbf{\bibinfo{volume}{16}},
  \bibinfo{pages}{6} (\bibinfo{year}{2013}), \eprint{1206.1225}.

\bibitem[{\citenamefont{Kogut et~al.}(2011)\citenamefont{Kogut, Fixsen, Chuss,
  Dotson, Dwek et~al.}}]{Kogut:2011xw}
\bibinfo{author}{\bibfnamefont{A.}~\bibnamefont{Kogut}},
  \bibinfo{author}{\bibfnamefont{D.}~\bibnamefont{Fixsen}},
  \bibinfo{author}{\bibfnamefont{D.}~\bibnamefont{Chuss}},
  \bibinfo{author}{\bibfnamefont{J.}~\bibnamefont{Dotson}},
  \bibinfo{author}{\bibfnamefont{E.}~\bibnamefont{Dwek}}, \bibnamefont{et~al.},
  \bibinfo{journal}{JCAP} \textbf{\bibinfo{volume}{1107}}, \bibinfo{pages}{025}
  (\bibinfo{year}{2011}), \eprint{1105.2044}.

\end{thebibliography}

\end{document}